\documentclass[conference]{IEEEtran}

\IEEEoverridecommandlockouts
\usepackage{cite}
\usepackage{float}
\usepackage{algorithm}
\usepackage{algpseudocode}
\usepackage{amsmath,amssymb,amsfonts}
\usepackage{graphicx}
\usepackage{textcomp}
\usepackage{multirow}
\usepackage[dvipsnames]{xcolor}
\usepackage{subcaption}
\usepackage[dvipsnames]{xcolor}

\def\BibTeX{{\rm B\kern-.05em{\sc i\kern-.025em b}\kern-.08em
    T\kern-.1667em\lower.7ex\hbox{E}\kern-.125emX}}

\begin{document}

\title{
Group Decision-Making System with Sentiment
Analysis of Discussion Chat and Fuzzy Consensus Modeling
}

\author{
    \IEEEauthorblockN{Adilet Yerkin}
    \IEEEauthorblockA{
        \textit{School of Information Technology and Engineering} \\
        \textit{Kazakh-British Technical University} \\
        Tole Bi street 59, Almaty, Kazakhstan\\
        \tt ad\_yerkin@kbtu.kz
    }
    \and
    \IEEEauthorblockN{Pakizar Shamoi}
    \IEEEauthorblockA{
        \textit{School of Information Technology and Engineering} \\
        \textit{Kazakh-British Technical University} \\
        Tole Bi street 59, Almaty, Kazakhstan\\
        \tt p.shamoi@kbtu.kz
    }
}

\maketitle

\begin{abstract}
Group Decision-Making (GDM) plays a crucial role in various real-life scenarios where individuals express their opinions in natural language rather than structured numerical values. Traditional GDM approaches often overlook the subjectivity and ambiguity present in human discussions, making it challenging to achieve a fair and consensus-driven decision. This paper proposes a fuzzy consensus-based group decision-making system that integrates sentiment and emotion analysis to extract preference values from textual inputs. The proposed framework combines explicit voting preferences with sentiment scores derived from chat discussions, which are then processed using a Fuzzy Inference System (FIS) to compute a total preference score for each alternative and determine the top-ranked option. To ensure fairness in group decision-making, we introduce a fuzzy logic-based consensus measurement model that evaluates participants' agreement and confidence levels to assess overall feedback. To illustrate the effectiveness of our approach, we apply the methodology to a restaurant selection scenario, where a group of individuals must decide on a dining option based on brief chat discussions.  The results demonstrate that the fuzzy consensus mechanism successfully aggregates individual preferences and ensures a balanced outcome that accurately reflects group sentiment.

\end{abstract}


\section{Introduction}

Decision-making is a fundamental part of human interactions, and group decision-making (GDM) frequently relies on natural language for communication. GDM \cite{Peniwati2017, 9306916} is a complex process that involves multiple individuals working together to reach a consensus or make a decision that satisfies everyone \cite{10284209}. This process is influenced by various factors, including group dynamics \cite{Perez2018On, Gupta2018Consensus}, individual biases \cite{2017Making, Bose2017Collective}, and external conditions. 

GDM models usually require participants to provide explicit numerical evaluations of options \cite{morente2018analysing}. However, in practical scenarios, people convey their preferences and reasoning through textual discussions, comments, and messages. This poses a challenge for computational systems, which must interpret subjective and context-dependent language. To address this issue, fuzzy natural language processing (NLP) techniques offer a promising approach by using sentiment analysis, fuzzy logic, and fuzzy set theory to process human language. Sentiment analysis enables the extraction of preference intensity from textual inputs. Meanwhile, fuzzy logic and fuzzy sets provide a framework to manage linguistic uncertainty and subjectivity in decision-making.

The mode of communication plays an important role in group decision-making. With advancements in technology, the dynamics of decision-making have evolved, especially with the rise of computer-mediated communication (CMC). Today, many decisions are made through online discussions via chats, emails, and other digital platforms \cite{Golzadeh2019OnTE, Sadovykh2015}, enabling more accessible collaboration.

Face-to-face communication allows for more interaction, leading to faster discussions and quicker consensus. On the other hand, CMC helps overcome physical and social barriers, making participation more equal among group members. However, CMC can also cause delays in decision-making and lower member satisfaction due to the lack of real-time responses and personal connections \cite{Kiesler1992Group, Baltes2002Computer-Mediated}. Unlike face-to-face discussions, CMC fails to capture emotions, which can be important in reaching a consensus. Achieving a consensus is essential to ensuring that all group members find the decision acceptable \cite{9253574,8413113,9306916}.

The current paper proposes a fuzzy consensus-based group decision-making methodology that integrates sentiment analysis to enhance fairness and achieve a balanced outcome reflecting group sentiment. The approach combines explicit voting preferences with sentiment scores extracted from textual discussions, which are processed using Fuzzy Logic, resulting in a total preference score. Next, we determine the top-ranked alternative, followed by an evaluation of a consensus using experts' agreement and confidence levels. The methodology is demonstrated through a restaurant selection scenario.

The study contributions are as follows:
\begin{itemize}
    \item \textit{Fuzzy-Based Consensus Model} – We develop a fuzzy logic-based consensus measurement model that evaluates agreement and confidence levels, ensuring a fair and human-consistent decision-making process.  
    \item \textit{Hybrid Preference Aggregation} – The system combines explicit voting preferences with sentiment scores and processes them through a Fuzzy Inference System (FIS) to compute a total preference score for each alternative. 
    \item \textit{Illustrative Application in Small Decision-Making} – We validate our approach with a restaurant selection scenario achieving high consensus level of human participants.
\end{itemize}

\section{Related Work}

\begin{figure}[tb]
    \centering
    \includegraphics[width=0.45\textwidth]{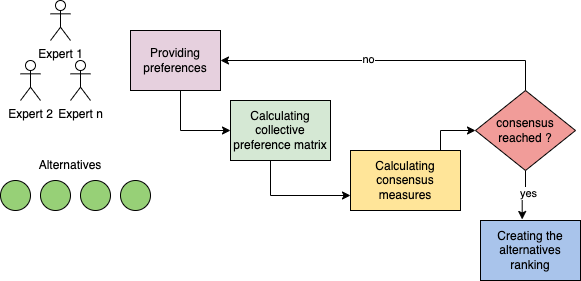}
    \caption{Basic process of a GDM system}
    \label{fig_1}
\end{figure}

GDM plays an important role in a number of fields, enabling multiple individuals to collaborate and reach a consensus on a shared decision. Fig. \ref{fig_1} provides a graphical representation of the group decision-making process. 

GDM includes two processes, namely, the selection process and consensus process \cite{1516155}. The selection process results in choosing the set of alternatives by aggregating expert opinions and analyzing preferences \cite{Cabrerizo2009}. The consensus process focuses on agreement among experts regarding the solution set of alternatives \cite{Alonso2009, kacprzyk2012consensus}. Traditional GDM systems rely on numerical preference aggregation and structured voting mechanisms. However, these approaches often face challenges such as communication barriers and conflicts \cite{2017Making, Bose2017Collective} and fail to accommodate the subjectivity and uncertainty present in real-world discussions \cite{Kacprzyk1992}. As a result, researchers have explored computational intelligence techniques, such as fuzzy logic and linguistic approaches, to enhance decision-making processes.

Fuzzy sets and logic help represent vague data, making it easier to capture group preferences and opinions \cite{Kacprzyk2008}. GDM systems use fuzzy sets and logic \cite{Zadeh1965, Zadeh1988} to support decision-making in uncertain or ambiguous situations \cite{Herrera2021}. Several studies explore the application of fuzzy theory in group GDM.

The fuzzy best-worst method (BWM) is a technique for GDM that combines individual and group decisions. It uses fuzzy preferences to reflect the imprecise nature of expert opinions and increases the consistency of group decisions \cite{best-worst1, best-worst2}. In large-scale group decision-making (LGDM), consensus models are essential for achieving agreement, and a novel approach utilizing dynamic clustering with hesitant fuzzy information and possibility distribution-based hesitant fuzzy elements (PDHFE) allows clusters to adapt as preferences evolve, improving the consensus process \cite{Wu2018A}. Fuzzy multi-criteria decision-making (MKDM) methods, like fuzzy AHP and fuzzy TOPSIS, handle complex multi-criteria decisions by reducing uncertainty and information loss, making them effective for tasks such as ERP system selection \cite{Akram2019Group, Efe2016An}. Intuitionistic and hesitant fuzzy sets help handle uncertainty in group decision-making by allowing both positive and negative opinions \cite{Meng2020Group, Alcantud2019Necessary}. They improve consensus processes by balancing expert preferences while achieving group agreement. GDM with incomplete fuzzy preference relations is explored in \cite{Herrera2007a}, where the consensus-reaching process is automated without a moderator, relying on consistency criteria.

People tend to express their opinions in natural language during discussions \cite{Herrera1996, Yerkin2024}. NLP enables experts to communicate using everyday language instead of structured formats or numerical values. Sentiment analysis \cite{Herrera2020a}, \cite{Trillo2022}, \cite{Herrera2019a} captures the overall sentiment tone of group members regarding an alternative classifying the opinion as positive, negative, or neutral or using the emotion classes like sad, happy, etc.

Several methodologies have been developed to integrate sentiment analysis into decision-making. The SA-MpMcDM approach uses deep learning to analyze expert reviews and aggregate evaluations for better decision support \cite{Zuheros2020Sentiment}. Other frameworks, like SentiRank, combine user preferences with public opinions to rank alternatives \cite{Jabreel2020Introducing}. A study \cite{Trillo2022} introduced a large-scale GDM method that utilizes sentiment analysis to process input from numerous experts.

Several studies proposed the integration of sentiment analysis with fuzzy logic to improve decision-making. For instance, fuzzy logic is used to handle ambiguity in sentiment analysis, providing better sentiment predictions by incorporating fuzziness with deep learning techniques like LSTM networks \cite{Bedi2019Sentiment}.

Fuzzy sentiment analysis is particularly useful in product ranking and decision-making.  Intuitionistic fuzzy sets (IFS) were used to represent customer reviews, capturing hesitant expressions and uncertainty in decision matrices \cite{Liu2017Ranking}. Another study \cite{Herrera2019a} proposed a GDM model that integrates free text inputs with pairwise comparisons of options. The other study proposed a consensus model for social network-based GDM \cite{Liu2023}. Recent studies include a fuzzy attention fusion-based Multimodal Sentiment Analysis (MSA) method \cite{10613477}, a hybrid approach combining sentiment analysis, fuzzy cognitive maps, and multi-criteria decision-making to improve product ranking from online customer reviews \cite{10313137}, tourism recommender systems \cite{Yerkin2024, 10478368}, e.g., combining the Artificial Bee Colony algorithm and Fuzzy TOPSIS to optimize travel recommendations  \cite{10478368}.

\section{Methodology}

Traditional group decision-making system that relies on the preference values of experts may be defined through several steps:
\begin{itemize}
  \item
Providing preferences: Experts will decide which alternatives are the most appropriate, therefore they should provide their preferences and opinions for a particular group of alternatives according to certain criteria. This evaluation is often done using scales, such as numerical ratings, numerical value, etc.
  \item
Calculating Collective Preference: The individual preferences are combined into a collective preference matrix, which represents the collective expert opinion of the alternatives, after all experts have expressed their preferences for the alternatives.
  \item
Creating the Alternatives Ranking: The final Rating of the Alternatives is calculated using the Collective Preference Matrix. The alternatives are ranked according to the preferences provided by experts.
  \item
Measuring consensus: The calculation of consensus measures enables us to determine whether experts have reached a consensus.  

\end{itemize}

 In many group decision-making systems \cite{delMoral2018, Enrique2002, Chiclana2013}, the calculation of consensus measures depends on assessing the similarity among experts’ opinions. This typically involves a coincidence or similarity function that determines the degree of agreement between the experts, identifying the degree of divergence of each expert from the group’s ones. However, these functions do not assess individual expert satisfaction or their agreement with the final decision. Instead, the system autonomously calculates the consensus level using expert preference data.

\begin{figure*}[tp]
    \centering
    \includegraphics[width=1\textwidth]{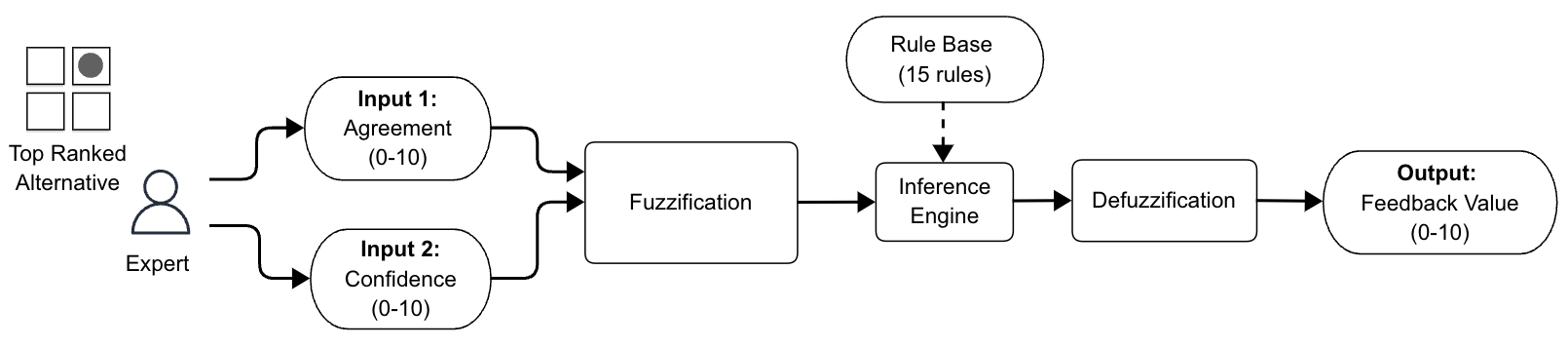}
    \caption{Consensus Estimation: Fuzzy Inference System}
    \label{fig_infer}
\end{figure*}

\begin{figure}[tp]
    \centering
    \includegraphics[width=0.45\textwidth]{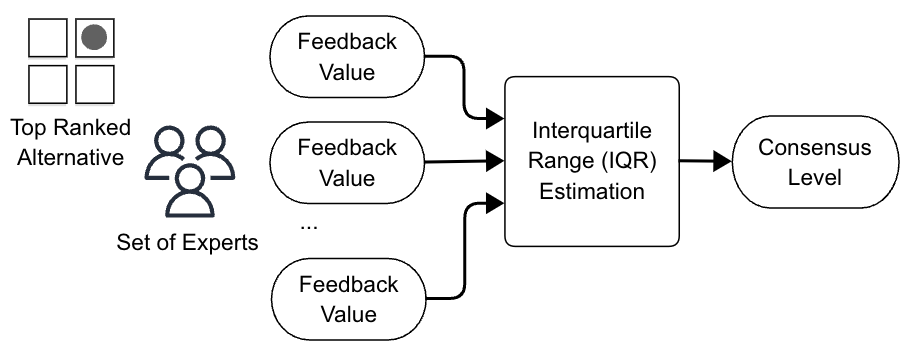}
    \caption{Human-consistent consensus measure}
    \label{fig_hccm}
\end{figure}

In our approach, we propose a human-consistent consensus measure that is based on the experts' agreement and their confidence level in the agreement in the system's outcome (see Fig. \ref{fig_hccm}). The consensus level of each expert is evaluated using a fuzzy inference system illustrated in Fig. \ref{fig_infer}. Algorithm 1 illustrates the proposed GDM System described in subsequent subsections.

\begin{algorithm}
\caption{Proposed Fuzzy GDM System}
\begin{algorithmic}[1]
    \State \textbf{Input:} Set of alternatives $X = \{x_1, x_2, \dots, x_n\}$, with a Set of Features $F = \{f_1, f_2, \dots, f_p\}$, Set of experts $E = \{e_1, e_2, \dots, e_m\}$
    \State \textbf{Output:} Top Ranked Alternative, Consensus Level

    \State \textbf{Step 1: Experts Provide Preferences}
    \For{each expert $e_j \in E$}
        \State Provide \textbf{Voting Preference} $V_{j,i}$ for each feature $f_i$
        \State Extract \textbf{Sentiment Preference} $S_{j,i}$ from chat text $(-1 \leq S_{j,i} \leq 1)$
    \EndFor

    \State \textbf{Step 2: Sentiment and Emotion Analysis}
    \For{each expert $e_j$}
        \State Extract \textbf{Sentiment Score} using VADER analysis: $S_j = f(\text{VADER}(T_j))$
        \State Extract \textbf{Emotion Score} using Text2Emotion: $E_j = g(\text{Emotion}(T_j))$
        \State Compute \textbf{Total Sentiment Preference}:  
            \[
            SP_j = \alpha S_j + \beta E_j
            \]
        \State Normalize sentiment scores
    \EndFor

    \State \textbf{Step 3: Compute Collective Preference}
    \For{each alternative $x_i$}
        \State Compute aggregated Voting Preference:
            \[
            V_{i} = \frac{1}{m} \sum_{j=1}^{m} V_{j,i}
            \]
        \State Compute aggregated Sentiment Preference:
            \[
            S_{i} = \frac{1}{m} \sum_{j=1}^{m} SP_{j,i}
            \]
    \EndFor

    \State \textbf{Step 4: Fuzzy Inference System  Processing}
    \For{each alternative $x_i$}
        \State Apply Fuzzy Logic with inputs: $V_{i}$ and $S_{i}$
        \State Compute \textbf{Total Preference Score} $T_{i}$ (range 0-10)
    \EndFor

    \State \textbf{Step 5: Rank Alternatives}
    \State Sort alternatives based on $T_{i}$ in descending order
    \State Select \textbf{Top Ranked Alternative} with the highest $T_{i}$

    \State \textbf{Step 6: Consensus Measurement}
    \For{each expert $e_j$}
        \State Collect \textbf{Agreement Level} ($A_j$) and \textbf{Confidence Level} ($C_j$)
        \State Compute \textbf{Feedback Score} using Fuzzy Inference
    \EndFor
    \State Compute \textbf{Interquartile Range (IQR)} for consensus evaluation
    \If{$IQR < 2.0$} 
        \State High consensus
    \ElsIf{$2.0 \leq IQR \leq 4.0$}
        \State Medium consensus
    \Else
        \State Low consensus
    \EndIf

    \State \textbf{Return:} Top Ranked Alternative and 
    Consensus Level
\end{algorithmic}
\end{algorithm}

\begin{figure*}[tb]
    \centering
    \includegraphics[width=0.75\textwidth]{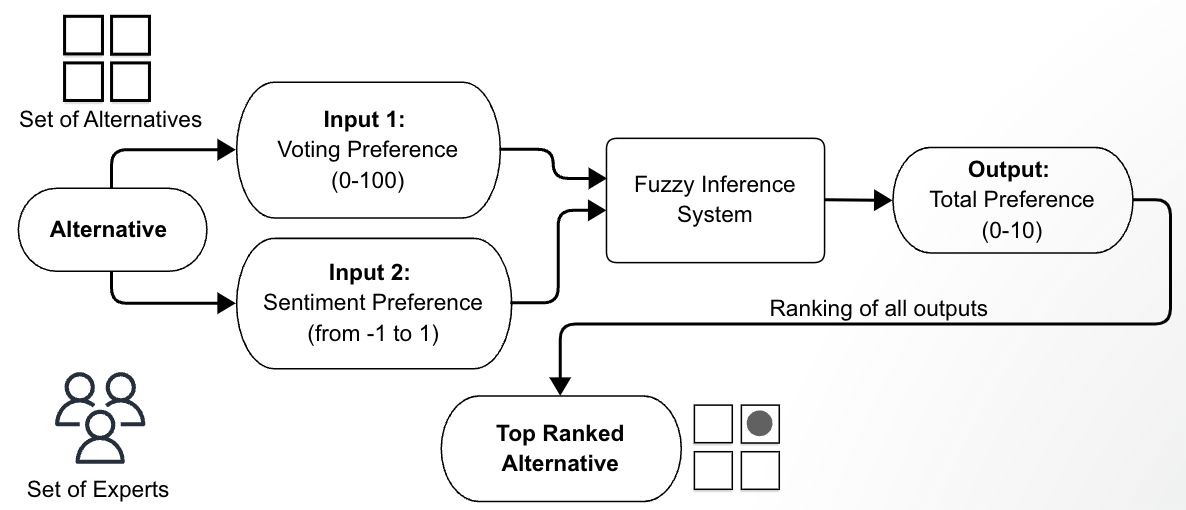}
    \caption{Integrating voting preference and sentiment preference to determine the total preference score and rank alternatives}
    \label{fig_pref_system}
\end{figure*} 

\subsection{Preferences, voting system}

Let $X = \{x_1, x_2,…, x_n\}$ be a set of alternatives, and let $E = \{e_1, e_2,..., e_m\}$ represent a finite set of experts tasked with evaluating these alternatives. Each alternative is described by a set of features $F = \{f_1, f_2, ..., f_p\}$, which are used to assess the alternatives in terms of predefined criteria.

Each expert $e_j$ provides a preference assessment, $Z = \{z_1, z_2,..., z_{p}\}$, for each feature, where the values
“-1” - against, “0” - does not matter, and “1” – agreement.
These assessments express each expert's opinion on the desirability of using a particular feature value in the evaluation of an alternative. This approach highlights the relative importance each expert assigns to the feature in question when assessing alternatives. In order to ensure that more relevant or knowledgeable expert opinions have a greater impact on the final decision, the weight of each expert $e_j$ can be adjusted to reflect their relative importance or expertise in the domain, depending on the particular requirements of the task.

GDM systems produce a ranked list of alternatives from the most preferred to the least preferred in $X$ by considering the preference values $Pref^{e_j}(x_i)$ given by each expert $e_j$, where $1 <= i <= n$ and $1 <= j <= m$.
$Pref^{e_j}(x_i)$ – preference value of expert, $e_j$, about alternative, $x_i$, calculated by using Eq. \ref{eqpref}.

\begin{equation}
Pref^{e_j}(x_i) = \sum_{i,j,k=1}^{i=n,j=m,k=p}f_k (x_i )*z_k (e_j)
\label{eqpref}
\end{equation}

\subsection{Sentiment and emotion analysis}

Through a chat messenger, experts debated the benefits and drawbacks of each option. This extensive but informal discussion provided insightful qualitative information. To extract sentiment and emotional scores from the conversation, we used text analysis tools. Using formula as $SP = \alpha S + \beta E$ , we aggregate the Sentiment and Emotion Scores to obtain a final sentiment preference score. Where $\alpha$ and $\beta$ are weights balancing the importance of sentiment and emotion.


The VADER sentiment analysis tool \cite{Hutto2014} from the NLTK library was used to get the sentiment score for every statement. This tool gives a compound score that ranges from -1 (negative) to 1 (positive).  The next step is to average the sentiment scores for each alternative after determining the sentiment scores for each expert. The Text2Emotion library \cite{t2e}, which evaluates various emotions, was used to identify the emotional tones (\textit{surprise, fear, sad, happy and angry }).
The positive and negative emotions were taken into consideration independently to determine the emotion score. It was determined using the maximum of the corresponding feelings: "happy" and "surprise" were examples of positive emotions, whereas "angry," "sad," and "fear" were examples of negative emotions. The combined emotion score was calculated using the difference between the positive and negative emotion scores. The combined emotion score was calculated as follows:
\begin{align*}
Positive Emot. Score(PES) = max('happy', 'surprise')  \\
Negative Emot. Score(NES) = max ('angry', 'sad', 'fear') \\
Emotion Score (combined) = PES - NES
\end{align*}
This method guarantees that emotions both positive and negative are taken into account, reflecting the discussion's overall emotional tone. These scores were combined to form the total sentiment score: 
$Total sentiment Score = Sentiment Score * 0.6 + Emotion Score * 0.4 $.

This weighting system takes emotional intensity into account while giving sentiment analysis greater weight. Because sentiment analysis has a wider contextual significance, these weights were used to give it a little more weight. This weighting system guarantees that emotional intensity is still heavily considered even though sentiment analysis takes priority. In principle, they can be modified according to the situation or depending on requirements.

In our approach, we use a combined measure of sentiment and emotion when there is sufficient textual discussion, as longer discussions provide richer emotional context. However, emotions are complex, so longer texts allow for the analysis of emotion progression over time, which is essential for understanding how they develop \cite{long_em}, \cite{long_em2}. In cases where discussions are short, relying solely on sentiment analysis is preferable since extracting emotional patterns from limited text can be challenging. Sentiment analysis remains effective in such scenarios as it captures the overall polarity of opinions, whereas emotion analysis requires a more extensive linguistic input. Therefore, the choice between a combined sentiment-emotion measure and pure sentiment analysis depends on the length and depth of textual discussions within the group decision-making process.

\subsection{Fuzzy Inference System for Preference calculation}

As can be seen in Fig. \ref{fig_pref_system}, the system processes a set of alternatives evaluated by experts,  providing two key inputs:

\begin{itemize}
    \item \textbf{Input 1: Voting Preference (0-100)} – A numerical score reflecting the explicit voting preference of each expert for a given alternative.
    \item \textbf{Input 2: Sentiment Preference (-1 to 1)} – A sentiment-based evaluation derived from textual discussions, where -1 represents strong disapproval, 0 is neutral, and 1 indicates strong approval.
\end{itemize}

These inputs are processed using a FIS, which combines them to produce an \textbf{Output: Total Preference (0-10)}. This final preference score is used to rank all alternatives. The alternative with the highest total preference score is selected as the Top Ranked Alternative.

\begin{figure*}[t!]
    \centering
    \begin{subfigure}{0.3\textwidth}
        \centering
        \includegraphics[width=\textwidth]{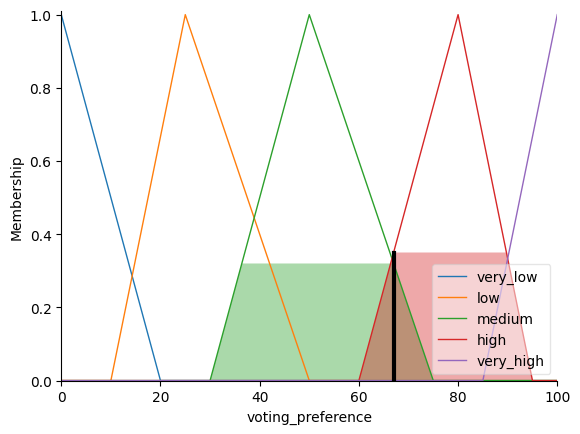}
        \caption{Voting pref-ces}
        \label{fi_voting_pref_fuz}
    \end{subfigure}
    \begin{subfigure}{0.3\textwidth}
        \centering
        \includegraphics[width=\textwidth]{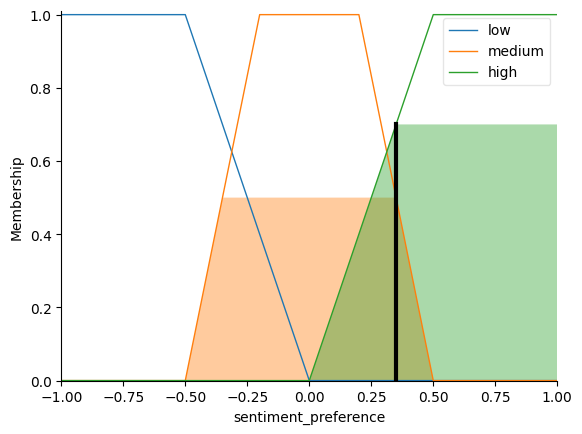}
        \caption{Sent. pref-ces}
        \label{fi_sentim_pref_fuz}
    \end{subfigure}
    \begin{subfigure}{0.3\textwidth}
        \centering
        \includegraphics[width=\textwidth]{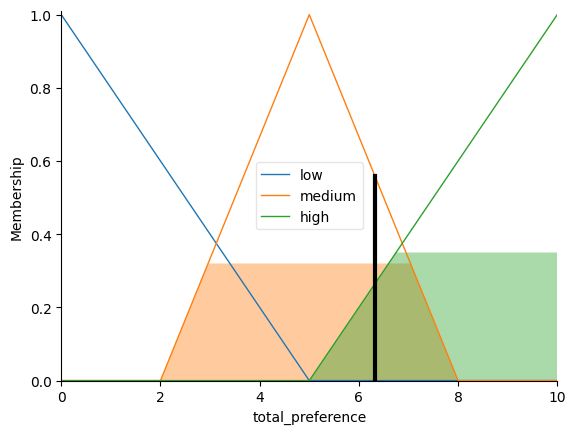}
        \caption{Total pref-ces}
        \label{fi_total_pref_fuz}
    \end{subfigure}
    \caption{FIS for Preference calculation}
    \label{fig:fi_pref_fuz}
\end{figure*}

We define a set of 15 fuzzy rules based on various combinations of Voting and Sentiment Preferences, each resulting in a Total Preference. 
An example of a fuzzy inference system (FIS) is shown in Figure \ref{fig:fi_pref_fuz}.

\subsection{Human-consistent consensus measure}

\subsubsection{Calculation Feedback score using Fuzzy inference system}
After Alternatives Ranking and selecting the best alternative, each expert $e_i$  give feedback an agreement levels, $al_i$, and confidence levels in the agreement, $cl_i$, regarding the decision. 
The set of agreement levels is denoted as $\mathcal{AL} = \{al_1, al_2, ..., al_n\}$, with with $al_i \in [0, 10]$, and the set of confidence levels as $\mathcal{CL} = \{cl_1, cl_2, ..., cl_n\}$, with $cl_i \in [0, 10]$ for all experts $i = 1, 2, ..., n$.

\begin{figure}[t!]
        \centering
        \includegraphics[width=0.4\textwidth]{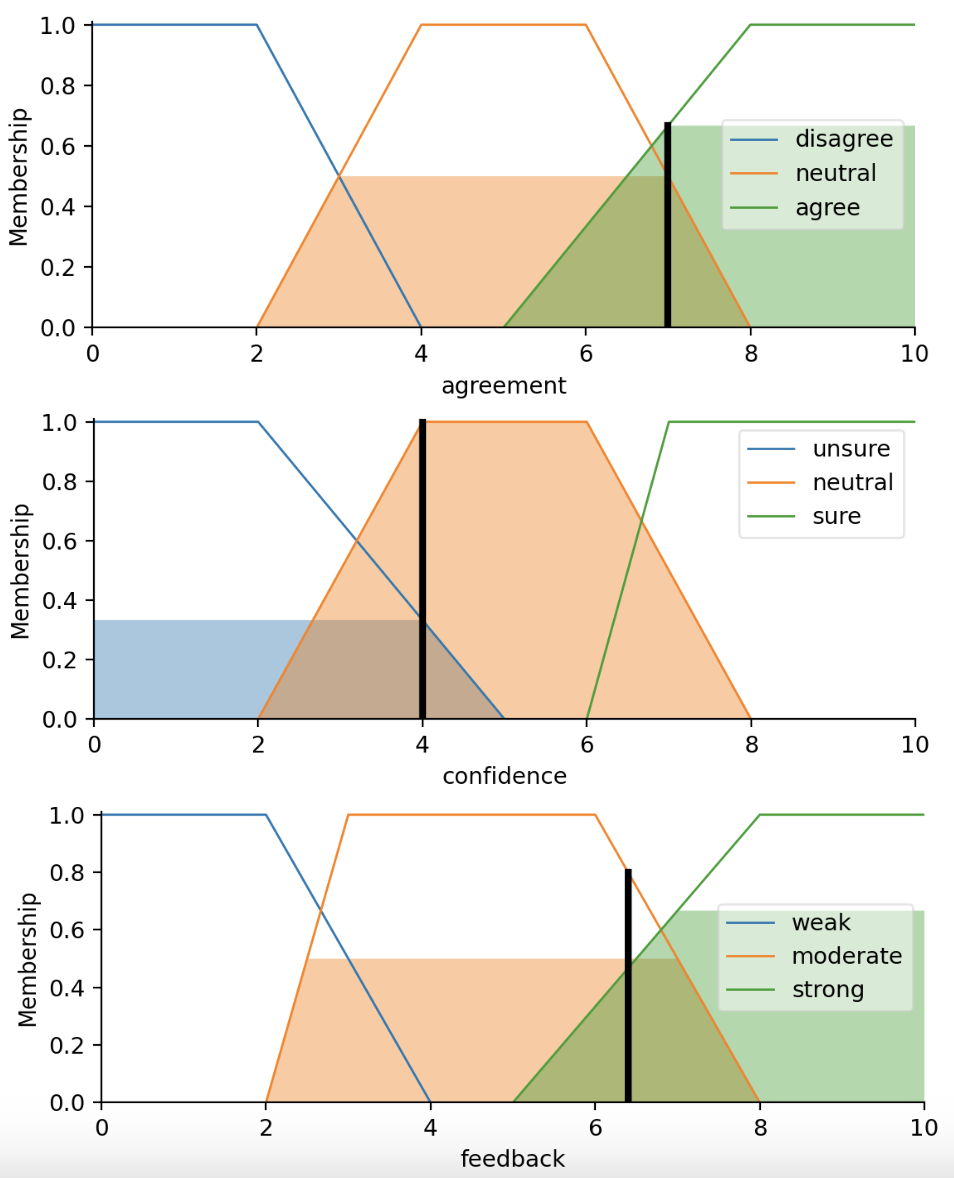}
        \caption{Example of feedback evaluation. the \textit{Agreement} = 7, and the \textit{Confidence} = 4. The resultant \textit{Feedback value} $\approx$ 6.4.}
        \label{fig11}
    \label{fig:combined_layout}
\end{figure}

We implemented a fuzzy inference system to calculate a total feedback score using the agreement and confidence inputs from experts (see Fig. \ref{fig11}). The universe of discourse for each fuzzy variable—agreement, confidence, and feedback—is distributed into a range from $[0, 10]$. For each input variable's agreement and confidence, we define several fuzzy sets representing linguistic terms. For example, agreement levels are categorized into sets such as \textit{A = $\{$'disagree', 'neutral', 'agree',  $\}$}, each represented by trapezoidal membership functions. Confidence levels are partitioned into \textit{C = $\{$ 'unsure', 'neutral', 'sure'$\}$}. 
The output variable, feedback, is also categorized into linguistic terms like \textit{T = $\{$ 'weak', 'moderate',  'strong'$\}$} , with corresponding fuzzy sets designed to represent the synthesized feedback based on the rules defined.

This system uses fuzzy logic to process inputs of experts’ agreement and confidence levels and outputs a feedback score.  The trapezoidal membership function is one of the standard shapes used in fuzzy logic systems due to its flexibility and simplicity. 
The general form of a trapezoidal membership function $\mu(x)$ can be described mathematically as: $\mu(x) = \max\left(\min\left(\frac{x - a}{b - a}, 1, \frac{d - x}{d - c}\right), 0\right)$


Fuzzy rules are formulated based on logical associations between the input and output sets. For instance, if an expert \textit{'agree'} and is \textit{'neutral'} in their confidence, the feedback is measured as \textit{'moderate'}. The rules are illustrated in Table \ref{tab:fuzzy_rules_feed}.

\begin{table}[]  
\caption{Fuzzy rules for feedback scores}
\label{tab:fuzzy_rules_feed}
\centering
\begin{tabular}{|c|c|c|}
\hline
\textbf{Agreement Level} & \textbf{Confidence Level} & \textbf{Feedback Score} \\ \hline
Agree                    & Unsure                    & Moderate                \\ \hline
Agree                    & Neutral                   & Moderate                \\ \hline
Agree                    & Sure                      & Strong                  \\ \hline
Neutral                  & Unsure                    & Moderate                \\ \hline
Neutral                  & Neutral                   & Moderate                \\ \hline
Neutral                  & Sure                      & Strong                  \\ \hline
Disagree                 & Unsure                    & Moderate                \\ \hline
Disagree                 & Neutral                   & Weak                    \\ \hline
Disagree                 & Sure                      & Weak                    \\ \hline
\end{tabular}
\end{table}

The linguistic terms of fuzzy variables and rules can be modified to suit different tasks or contexts. This adaptability ensures that the system remains relevant in different decision-making scenarios, allowing it to be customized based on the specific characteristics and requirements of each task.

\subsubsection{Calculation Consensus level}
 To evaluate the overall consensus among experts in the GDM process, we employ the interquartile range (IQR). It is a measure of statistical dispersion, which is especially helpful in determining the distribution of consensus scores within the group. Calculated as $IQR = Q_{3} - Q_{1}$, where $Q_{1}$ is the first quartile, $Q_{3}$ is the third quartile of distribution. Using a measurable and understandable metric such as IQR can increase the transparency of the group decision-making process. The thresholds for classifying consensus levels (\textit{high, medium}, or\textit{ low}) can be adjusted based on the context or importance of the decision, allowing flexibility in how consensus is evaluated and interpreted, for example, shown in Table \ref{consensus_level_meth}.


\begin{table}[tb]
    \caption{Consensus level}
    \centering
    \begin{tabular}{|c|l|}
    \hline
    \textbf{\begin{tabular}[c]{@{}c@{}}IQR\end{tabular}} & \multicolumn{1}{c|}{\textbf{Consensus level}} \\ \hline
    0.00 - 2.00                                                                   & High                                          \\ \hline
    2.01 - 4.00                                                                   & Medium                                        \\ \hline
    \textgreater{}= 4.01                                                          & None                                          \\ \hline
    \end{tabular}
    \label{consensus_level_meth}
\end{table}

\section{Numerical Example - Sample Application}

In our study, we illustrate the proposed decision-making framework through a numerical example of a small-scale decision—choosing a restaurant for dinner. Since this type of decision is typically made quickly and involves brief discussions, we rely solely on sentiment analysis rather than a combined sentiment-emotion measure.

The process begins with a group of friends considering several restaurant options and expressing their preferences through messages in a group chat. Sentiment analysis is applied to extract the overall positive, negative, or neutral stance from these textual inputs. Each restaurant is then assigned a sentiment preference score, which, along with explicit voting scores, is fed into a fuzzy inference system (FIS) to compute the total preference score for each alternative. The restaurant with the highest total preference score is selected as the most suitable choice. Then, the consensus is evaluated.

\begin{figure}[tp]
    \centering
    \includegraphics[width=0.42\textwidth]{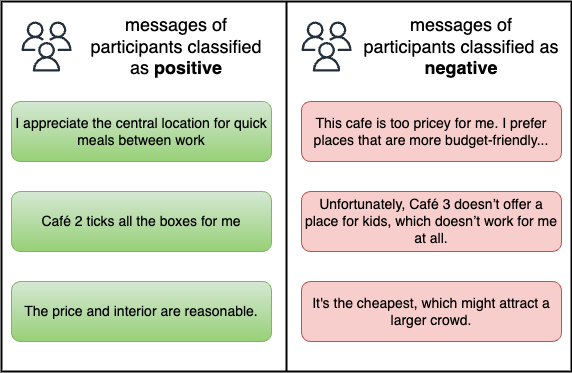}
    \caption{Examples of messages from chat discussions in the restaurant selection experiment scored as positive and negative.}
    \label{fig_pos_neg}
\end{figure}

In the voting part, the data comprises two components: 
\begin{itemize}
\item Alternative Characteristics: Each restaurant is characterized by several features: $feat_1$ - affordability (measured in monetary terms, average restaurant bill), $feat_2$ - location (rating - how much further from the city center), $feat_3$ - availability of vegan options (binary indicator), $feat_4$ - availability of child-friendly areas (binary indicator), and $feat_5$ - interior design (average rating from organizers), as shown in Table \ref{tab:cafes}.
\item Participant Preferences: Each participant provides preference assessments on a set of features for restaurants, as shown in Table \ref{tab:pref_assess}. 

\end{itemize}

\begin{table}[tb]
\caption{Features of alternatives}
\label{tab:cafes}
 \centering
\begin{tabular}{|l|l|l|l|l|l|}
\hline
                & \textbf{feat1} & \textbf{feat2} & \textbf{feat3} & \textbf{feat4} & \textbf{feat5} \\ \hline
\textbf{alter1} & 7500           & 1              & 1              & 0              & 3              \\ \hline
\textbf{alter2} & 9000           & 2              & 1              & 1              & 5              \\ \hline
\textbf{alter3} & 4000           & 2              & 0              & 0              & 2              \\ \hline
\textbf{alter4} & 8000           & 3              & 0              & 0              & 4              \\ \hline
\end{tabular}
\end{table}

\begin{table}[]
\caption{Preference assessment of each expert}
\label{tab:pref_assess}
 \centering
\begin{tabular}{|c|r|r|r|r|r|}
\hline
\multicolumn{1}{|l|}{} &
  \multicolumn{1}{l|}{\textbf{feat1}} &
  \multicolumn{1}{l|}{\textbf{feat2}} &
  \multicolumn{1}{l|}{\textbf{feat3}} &
  \multicolumn{1}{l|}{\textbf{feat4}} &
  \multicolumn{1}{l|}{\textbf{feat5}} \\ \hline
\textbf{partp1} & 0 & 0 & 1  & 1  & 1 \\ \hline
\textbf{partp2} & 0 & 1 & 1  & 0  & 0 \\ \hline
\textbf{partp3} & 1 & 1 & 1  & 0  & 1 \\ \hline
\textbf{partp4} & 1 & 0 & -1 & 1  & 0 \\ \hline
\textbf{partp5} & 0 & 1 & 0  & -1 & 1 \\ \hline
\end{tabular}
\end{table}

We find the participants' preference scores regarding each restaurant using the following stages:
\begin{itemize}
\item Normalization of Continuous Features: Continuous features such as affordability, location, and interior design are normalized. 

\item Affordability: If the restaurant's affordability is less than the mean affordability, it is considered affordable (boolean value of 1). Otherwise, it is not  (0).
\item Location: A location rating below the mean is considered less favorable (boolean value of 1), and above the mean is considered favorable (boolean value of 0).
\item Interior Design: A rating above the mean is deemed favorable (boolean value of 1), and below the mean as less favorable (boolean value of 0).
\item Direct Use of Binary Features: For features already in binary form, such as the availability of vegan options and child-friendly areas, the values are directly used in the preference score calculation without normalization.
\end{itemize}

The group preference value is determined by averaging the preference values provided by the experts (see Table \ref{tab:pref_assess_matrix_2}).

\begin{table}[]
\caption{Participants preference matrix based on voting part}
\label{tab:pref_assess_matrix_2}
\centering
\begin{tabular}{|l|ccccc|ccccc|c|}
\hline
\textbf{} &
  \multicolumn{5}{c|}{\textbf{\begin{tabular}[c]{@{}c@{}}Preference value\end{tabular}}} &
  \multicolumn{5}{c|}{\textbf{\begin{tabular}[c]{@{}c@{}}Scaled pref. values 
  \end{tabular}}} &
  \multirow{2}{*}{\textbf{\begin{tabular}[c]{@{}c@{}}Avg. \\ Score\end{tabular}}} \\ \cline{1-11}
\textbf{partID} &
  \multicolumn{1}{c|}{\textbf{1}} &
  \multicolumn{1}{c|}{\textbf{2}} &
  \multicolumn{1}{c|}{\textbf{3}} &
  \multicolumn{1}{c|}{\textbf{4}} &
  \textbf{5} &
  \multicolumn{1}{c|}{\textbf{1}} &
  \multicolumn{1}{c|}{\textbf{2}} &
  \multicolumn{1}{c|}{\textbf{3}} &
  \multicolumn{1}{c|}{\textbf{4}} &
  \textbf{5} &
   \\ \hline
\textbf{alter1} &
  \multicolumn{1}{c|}{1} &
  \multicolumn{1}{c|}{1} &
  \multicolumn{1}{c|}{1} &
  \multicolumn{1}{c|}{-1} &
  0 &
  \multicolumn{1}{c|}{60} &
  \multicolumn{1}{c|}{60} &
  \multicolumn{1}{c|}{60} &
  \multicolumn{1}{c|}{40} &
  50 &
  \textbf{54} \\ \hline
\textbf{alter2} &
  \multicolumn{1}{c|}{3} &
  \multicolumn{1}{c|}{1} &
  \multicolumn{1}{c|}{2} &
  \multicolumn{1}{c|}{0} &
  0 &
  \multicolumn{1}{c|}{80} &
  \multicolumn{1}{c|}{60} &
  \multicolumn{1}{c|}{70} &
  \multicolumn{1}{c|}{50} &
  50 &
  \textbf{62} \\ \hline
\textbf{alter3} &
  \multicolumn{1}{c|}{0} &
  \multicolumn{1}{c|}{0} &
  \multicolumn{1}{c|}{1} &
  \multicolumn{1}{c|}{1} &
  0 &
  \multicolumn{1}{c|}{50} &
  \multicolumn{1}{c|}{50} &
  \multicolumn{1}{c|}{60} &
  \multicolumn{1}{c|}{60} &
  50 &
  \textbf{54} \\ \hline
\textbf{alter4} &
  \multicolumn{1}{c|}{1} &
  \multicolumn{1}{c|}{1} &
  \multicolumn{1}{c|}{2} &
  \multicolumn{1}{c|}{0} &
  2 &
  \multicolumn{1}{c|}{60} &
  \multicolumn{1}{c|}{60} &
  \multicolumn{1}{c|}{70} &
  \multicolumn{1}{c|}{50} &
  70 &
  \textbf{62} \\ \hline
\end{tabular}
\end{table}

Fig. \ref{fig_pos_neg} shows examples of messages from the collected dataset with chat discussion. For each restaurant in chat messenger, the VADER sentiment analyzer provides a compound sentiment score. The compound score is a single metric that ranging from -1 (extremely negative) to +1 (extremely positive).
Each participant's comment about a restaurant is processed to calculate sentiment scores, as shown in Table \ref{tab:Sentiment_Preference_Score}. 
As seen in Table \ref{tab:Sentiment_Preference_Score}, the sentiment scores for each restaurant are then averaged to determine the average sentiment score for each restaurant.

\begin{table}[tb]
\caption{Sentiment preference matrix of participants}
\label{tab:Sentiment_Preference_Score}
\centering
\begin{tabular}{|l|c|c|c|c|c|c|}
\hline
\multicolumn{1}{|c|}{\textbf{}} &
  \textbf{parp1} &
  \textbf{parp2} &
  \textbf{parp3} &
  \textbf{parp4} &
  \textbf{parp5} &
  \textbf{\begin{tabular}[c]{@{}c@{}}avg. \\ score\end{tabular}} \\ \hline
\textbf{alter1} & 0.42  & 0.00 & 0.00 & 0.70 & -0.07 & \textbf{0.21} \\ \hline
\textbf{alter2} & 0.81  & 0.95 & 0.00 & 0.80 & 0.78  & \textbf{0.67} \\ \hline
\textbf{alter3} & 0.06  & 0.36 & 0.34 & 0.77 & 0.51  & \textbf{0.41} \\ \hline
\textbf{alter4} & -0.30 & 0.67 & 0.66 & 0.94 & 0.71  & \textbf{0.54} \\ \hline
\end{tabular}
\end{table}

After that, using the input values such as voting preference and sentiment preference (see Fig. \ref{fig:fi_pref_fuz}), we simulated and calculated the total preference score using the FIS. The total preference values are displayed in Table \ref{tab:total_pref_fuzzy}. 

\begin{table}[tb]
\caption{Total preference scores}
\label{tab:total_pref_fuzzy}
\centering
\begin{tabular}{|c|c|c|c|c|}
\hline
                                                                              & \textbf{alter1} & \textbf{alter2} & \textbf{alter3} & \textbf{alter4} \\ \hline
\textbf{\begin{tabular}[c]{@{}c@{}}Score\end{tabular}} & 5               & 5.99            & 5               & 5.36            \\ \hline
\end{tabular}
\end{table}

After the restaurant selection, participants provided their degree of agreement and confidence in the decision. To calculate a feedback score (see Fig. \ref{fig:combined_layout}), these values were calculated using a fuzzy inference system. As indicated in Table \ref{tab:tab_feedback}, the feedback system assessed the participants' degree of agreement and confidence. The consensus metric, which is the interquartile range (IQR) computed to assess the general level of agreement among participants. Based on the consensus score, the consensus level was categorized as \textit{high} (IQR=0.19), while the average feedback score among the participants was $\approx$ 8.1.

\begin{table}[tb]
\caption{Participants Feedback}
\label{tab:tab_feedback}
\centering
\begin{tabular}{|c|c|c|c|}
\hline
\textbf{\begin{tabular}[c]{@{}c@{}}id of  part.\end{tabular}} &
  \textbf{\begin{tabular}[c]{@{}c@{}}Agreement(0-10)\end{tabular}} &
  \textbf{\begin{tabular}[c]{@{}c@{}}Confidence(0-10)\end{tabular}} &
  \textbf{\begin{tabular}[c]{@{}c@{}}Feedback value\end{tabular}} \\ \hline
1 & 9  & 9  & 8.14 \\ \hline
2 & 10 & 10 & 8.14 \\ \hline
3 & 7  & 8  & 7.95 \\ \hline
4 & 9  & 9  & 8.14 \\ \hline
5 & 7  & 9  & 7.95 \\ \hline
\end{tabular}
\end{table}

\section{Conclusion}
This paper presented a fuzzy consensus-based group decision-making system that integrates sentiment analysis to enhance collective preference and account for experts' emotions. Our approach successfully combined explicit voting preferences with sentiment scores and processed them using an FIS to determine a total preference score for each alternative. The methodology was applied to a restaurant selection scenario, demonstrating its effectiveness in small-scale decision-making. The results showed that our model achieved a high level of consensus (IQR=0.19), ensuring that the final decision reflected the collective preference of the group.

The proposed system has limitations. The approach relies on the quality and quantity of textual discussions, so it may not perform optimally when chat data is limited.  For future work, we plan to extend the framework to incorporate adaptive weighting mechanisms based on context-specific factors. Furthermore, testing the system on larger decision-making tasks will help validate it in more complex settings.

\section*{Acknowledgment}
This research has been funded by the Science Committee of the Ministry of Science and Higher Education of the Republic of Kazakhstan (Grant No. AP22786412)

\bibliographystyle{IEEEtran}
\bibliography{export}
\end{document}